# An Empirical Analysis of Secure Federated Learning for Autonomous Vehicle Applications

Md Jueal Mia[1,2] and M. Hadi Amini[1,2]

[1]Knight Foundation School of Computing and Information Sciences, Florida International University, Miami, FL 33199; e-mails: {mmia001,moamini}@fiu.edu

[2] Sustainability, Optimization, and Learning for InterDependent networks laboratory (solid lab)

**ABSTRACT**

Federated Learning lends itself as a promising paradigm in enabling distributed learning for autonomous vehicles applications and ensuring data privacy while enhancing and refining predictive model performance through collaborative training on edge client vehicles. However, it remains vulnerable to various categories of cyber-attacks, necessitating more robust security measures to effectively mitigate potential threats. Poisoning attacks and inference attacks are commonly initiated within the federated learning environment to compromise secure system performance. Secure aggregation can limit the disclosure of sensitive information from outsider and insider attackers of the federated learning environment. In this study, our aim is to conduct an empirical analysis on the transportation image dataset (e.g., LISA traffic light) using various secure aggregation techniques and multiparty computation in the presence of diverse categories of cyber-attacks. Multiparty computation serves as a state-of-the-art security mechanism, offering standard privacy for secure aggregation of edge autonomous vehicles local model updates through various security protocols. The presence of adversaries can mislead the autonomous vehicle learning model, leading to the misclassification of traffic lights, and resulting in detrimental impacts. This empirical study explores the resilience of various secure federated learning aggregation techniques and multiparty computation in safeguarding autonomous vehicle applications against various cyber threats during both training and inference times.

**INTRODUCTION**

Autonomous Vehicles (AVs) are intelligent vehicles intended to replace human control. The Society of Automotive Engineers (SAE) classifies vehicle automation using a scale ranging from level 0 indicating no automation to level 5 representing full automation (SAE On-Road Automated Vehicle Standards Committee 2014). The software and automotive industries have recently unveiled a prototype of a self-driving car, which is expected to be available in the market soon (Fagnant and Kockelman 2015). AVs have various intelligent features or applications such as identifying, categorizing, and precisely locating objects, steering control, and mapping. Among these applications, we will focus on object recognition (e.g., traffic sign recognition) for this empirical study. Object recognition plays a pivotal role in helping AVs understand the driving environments. However, Technological advancements introduce various adversarial attacks in AV application development, necessitating robust security and privacy measures. Privacy breaches can result in severe consequences like public safety risks, authentication concerns, service disruptions, and communication vulnerabilities (Qayyum et al. 2020). Machine learning and deep learning are widely used techniques for the development of AV applications. However, their effectiveness in protecting data privacy has been undermined by security issues and data transmission costs associated with traditional data collection procedures. To address this concern, researchers have developed a federated learning (FL) (McMahan et al. 2017) framework aimed at preserving the privacy of individual user data. In this approach, data remains private on the edge client devices and collaborates with the central server by sharing model updates. The FL server lacks control over the edge client devices, rendering FL susceptible to various types of cyberattacks, including poisoning and inference attacks.

In poisoning attacks, an attacker can manipulate data or model parameters, which are then transmitted to the aggregator to disrupt the training process (Fang et al. 2020). These attacks can occur in the forms of data and model poisoning. Several studies have been implemented to address the adverse impacts of poisoning attacks, which are considerably more common in the practical context of FL environments. In inference attacks, the attacker can be an insider or outsider of the FL environment. In this scenario, a malicious attacker compromises the central model aggregator, compromising the model's integrity to deduce sensitive information about the data on edge client's devices through analysis of the aggregation results (Nasr et al. 2018). There are other categories of cyber-attacks, but these are the more common types observed in FL.

Several studies have been conducted to preserve the privacy of AVs. One prominent work introduced by (Chen et al. 2019) is the BDFL, a decentralized FL method that utilizes Byzantine-Fault-Tolerant (BFT) techniques and Publicly Verifiable Secret Sharing (PVSS) for secure model training. Evaluation on MNIST and KITTI datasets demonstrates BDFL's effectiveness in enhancing AV performance, particularly in multi-object recognition tasks, while safeguarding data privacy. (Olowononi et al. 2021) proposed privacy-aware FL with differential privacy for Vehicular Cyber-Physical Systems (VCPS), integrating technologies like SDN, blockchain, and ML, with FL operating at the edge for enhanced efficiency and security. The Optimized Quantum-based Federated Learning (OQFL) (Yamany et al. 2021) framework designed to safeguard against malicious attacks in AVs by employing quantum-behaved optimization on benchmark datasets, showcasing increased resilience and robustness. Moreover, SemBroc-RF (Zhu et al. 2023) handled data privacy and security challenges in intelligent transportation systems using reinforcement learning and partially encrypted multi-party computation, showcasing heightened accuracy and reduced training rounds. Finally, (Parekh et al. 2023) proposed GeFL that mitigated privacy concerns in AVs by refining local models and encrypting input data on edge devices, exhibiting improved accuracy and decreased data transmission among network devices compared to traditional FL methods.

All existing systems in AV are designed with specific cyber-attack resilient approaches. However, in real-time scenarios, multiple categories of cyber-attacks can occur simultaneously. Furthermore, while secure aggregation can provide security against poisoning attacks, if the aggregator is compromised, it can manipulate the entire FL system. Multiparty computation (MPC) can aggregate all the model parameters through a secure protocol to defend against inference attacks. SAFEFL (Gehlhar et al. 2023) has already implemented secure FL techniques with MPC using different protocols. We will use the same methods analyzed in the SAFEFL study with hyperparameter optimization on transportation datasets. Therefore, this study aims to evaluate the resilience of various aggregation techniques against seven categories of poisoning attacks and inference attacks in AV application. The main contributions of this study are outlined as follows.

1. We apply the methodologies examined in the SAFEFL study (Gehlhar et al. 2023) and tailored by hyperparameter optimization to transportation datasets for AV application development, considering seven categories of poisoning attacks and inference attacks.
2. The empirical analysis aims to find the robust, cyber-attack resilient secure aggregation techniques during FL training phase.
3. Furthermore, an in-depth analysis of the results has been conducted, focusing on influential factors and characteristics of adversarial attacks across multiple communication rounds. This research is set to establish a solid foundation for the development of AV applications.

**METHODOLOGY**

Our study involves implementing secure FL aggregation algorithms and Secure MPC to address various cyber-attacks. Figure 1 presents an overview of our empirical study, focusing on evaluating the resiliency of secure FL techniques in the context of AV applications, considering seven categories of poisoning attacks and inference attacks.

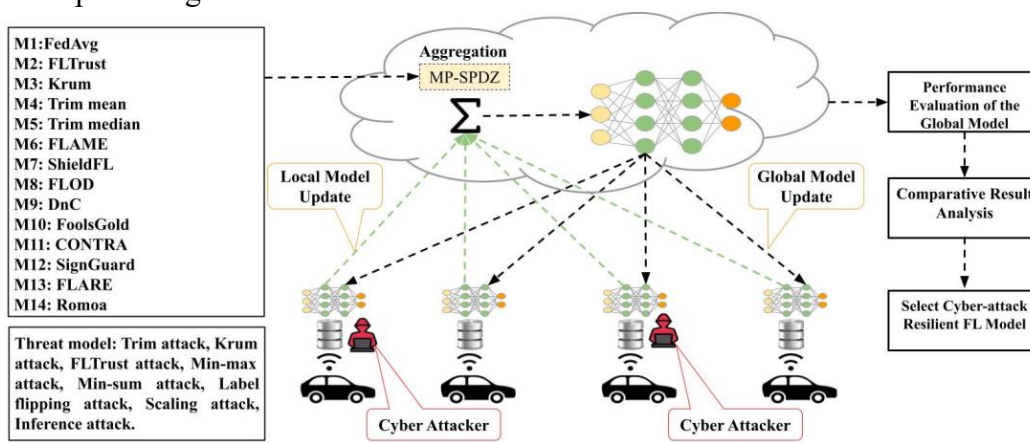

**Figure 1. Overview of the empirical analysis of secure FL for AV applications.**

In our flowchart, AVs serve as edge client devices, each having a private dataset. AVs receive the global model from the central server and refine it to enhance performance by sharing locally updated model parameters, instead of decentralizing data. Initially, we are particularly focused on evaluating the performance of different aggregation techniques in the presence of poisoning attacks. (Gehlhar et al. 2023) designed an MPC-based framework named SAFEFL that includes a communicator interface to connect two modules, namely model training, and aggregation. Model training takes place on the edge client device using PyTorch, while aggregation is performed using different MPC protocols of MP-SPDZ (Keller 2020). The main aim is to create robust, cyber-attack-resilient FL techniques that can mitigate poisoning and inference attacks. This approach is being evaluated in the context of AV applications. This analysis aims to identify a resilient model against cyber-attacks for AV traffic light detection applications.

**Threat model.** In this study, our threat model primarily focuses on poisoning attacks and inference attacks within the FL framework. For poisoning attacks, the attacker requires a compromised device to initiate data poisoning or model poisoning attacks. They can achieve this by manipulating either the training data within the compromised edge client device or by manipulating the model gradients/updates. These attacks aim to manipulate with the global model, leading to incorrect classifications. Another potential threat is inference attacks, where an attacker attempts to extract information or reveal data by analyzing the aggregated model updates. We will be considering the following seven categories of poisoning attacks and inference attacks previously introduced in various research studies.

**Trim attack** (Fang et al. 2020)**.** The Trim attack manipulate the gradients from selected malicious clients to disrupt the central server's aggregation process, exploiting vulnerabilities in the aggregation mechanism within FL systems.

**Krum attack** (Fang et al. 2020)**.** The Krum attack employs meticulous calculations based on Euclidean distances and iterative adjustments using lambda (scaling factor) values, aiming to disrupt the aggregation process.

**FLTrust attack** (Fang et al. 2020)**.** The FLTrust attack manipulates gradients from specified malicious clients. It employs Euclidean norms, weighted calculations, and iterative adjustments with parameters like gamma (scaling factor), eta (adjustment of gradient), and standard deviation, it compromises the integrity of the aggregated model.

**Min-Max and Min-Sum attack** (Shejwalkar and Houmansadr 2021)**.** The Min-max attack manipulates specific client's gradients, optimizing the gamma parameter to scale adversarie's mean gradient's deviation vector. By maximizing separation from other gradients, it disrupts FL's aggregation, compromising the model's integrity. Similarly, the Min-Sum attack computes mean gradients and utilizes gamma to emphasize differences, posing a threat to FL model reliability and security through aggregation manipulation.

**Label flipping attack** (Tolpegin et al. 2020)**.** This data poisoning attack changes labels on the training data of malicious clients by flipping them to their complement relative to the highest label number. Specifically targeted at the malicious clients, this attack aims to corrupt their labeled data, potentially causing misclassification or erroneous model learning during training.

**Scaling attack** (Wu et al. 2020)**.** This attack consists of two stages affecting training data and gradients contributed by compromised clients in FL. Initially, it inserts backdoor patterns into specific malicious client's training data, strategically modifying instances to embed trigger patterns, and adjusting labels to a predefined target. Subsequently, the attack amplifies the scale of gradients provided by these compromised clients, aiming to disrupt the learning process and compromise the integrity of FL models across diverse datasets.

**Inference attack** (Nasr et al. 2018)**.** An honest but curious or dishonest client can extract sensitive information by analyzing either the aggregated model or local model during inference time. Inference attacks are not detectable, so we can make our system resilient against attackers using MPC through secure aggregation.

**Aggregation techniques**

**M1.** Fedavg (McMahan et al. 2017) aggregates gradients from multiple client devices, computing a global model update by averaging these gradients weighted by each client's data size. The resulting global update is utilized to adjust the parameters of the central model using a specified learning rate.

**M2.** FLTrust (Cao et al. 2020) is a FL method that computes trust scores based on gradient similarities between clients and a server update, then aggregates these gradients to update the global model, giving more weight to trustworthy clients while handling potential byzantine attacks.

**M3.** Krum (Blanchard et al. 2017) detects and mitigates potential outlier or malicious gradients among participating clients by computing pairwise distances. It selects the global update based on the model with the smallest sum of distances to trustworthy models (excluding potential outliers) and updates the global model parameters accordingly.

**M4.** Trim mean (Yin et al. 2018) is a FL algorithm that mitigates potential outlier effects in gradients by trimming the largest and smallest values from the concatenated gradients of participating clients. It computes the global update by taking the mean of the remaining values and uses this update to adjust the global model parameters with a specified learning rate.

**M5.** Trim Median (Yin et al. 2018) is a FL method that computes the global update by determining the median of concatenated gradients from participating clients. This computed median is then used to update the global model parameters with a specified learning rate.

**M6.** FLAME (Nguyen et al. 2022) uses differential privacy and clustering to handle byzantine attacks. It computes cosine distances between gradients, utilizes HDBSCAN clustering for gradient grouping, applies gradient clipping based on clustering, adds adaptive noise for privacy, and finally updates the global model parameters using the aggregated and perturbed gradients.

**M7.** The ShieldFL (Ma et al. 2022) incorporates normalization of gradients, identifying outliers through cosine similarity, establishing a poison baseline gradient, computing confidence scores, and aggregating gradients for global model updates. It returns the updated global model and local model gradients for the subsequent iteration.

**M8.** The FLOD (Dong et al. 2021) method in FL employs gradient encodings, Hamming distances, and tau-clipping weights to aggregate gradients. It aims to mitigate the influence of gradients and updates global model parameters using a learning rate.

**M9.** Divide-and-Conquer (DnC) (Shejwalkar and Houmansadr 2021) iteratively selects random dimensions of the gradients, performs outlier detection using Singular Value Decomposition (SVD), and identifies reliable gradients. The reliable gradient's mean is computed for a global update, which is then used to update the global model parameters with a specified learning rate.

**M10.** FoolsGold (Fung et al. 2020) utilizes historical gradients to assign weights to current gradients based on cosine similarity. These weighted gradients constitute a global update that modifies global model parameters. It utilizes individual learning rates determined by dynamic scaling factors derived from cosine similarities.

**M11.** CONTRA (Awan et al. 2021) selects a fraction of clients according to reputation scores, aggregates their gradients using cosine similarity, and iteratively adjusts reputations and similarities. It computes individual learning rates and updates global model parameters with the computed global update.

**M12.** The SignGuard (Xu et al. 2021) technique in FL leverages sign statistics and clustering to detect malicious gradients. It filters based on gradient norms, calculates sign statistics, conducts sign-based clustering, selects pertinent gradients, and updates the global model accordingly.

**M13.** The FLARE (Wang et al. 2022) employs a process based on maximum mean discrepancy (MMD) between penultimate layer representations (PLR) of client's models on auxiliary data. It selects neighbors based on MMD values, counts their occurrences, and then calculates a global update using these counts. The resultant update is applied to the global model.

**M14.** The Romoa (Mao et al. 2021) is a FL technique that considers various distances between gradients and calculates a sanitization factor for aggregation. It selects parameters based on different distance metrics, including cosine similarity and Pearson correlation distance. The resulting sanitization factor is employed to update the global model.

**EXPERIMENTS AND RESULT ANALYSIS**
Our experiments were conducted on a machine equipped with 12th Gen Intel(R) Core (TM) i7-12650H 2.30 GH and 16 GB of RAM, running on Linux. We are utilizing the LISA real-time traffic light dataset, which includes red, green, and yellow traffic lights. This dataset is processed from real-time video. Initially, the dataset was distributed among edge client's devices, considering both IID and Non-IID data distributions. The evaluated results were based on the non-IID distribution of the data. We employed a CNN model consisting of two convolutional layers, batch normalization, and ReLU activation functions, the FL training was executed for 200 rounds of iterations. Various parameters were utilized for distinct aggregation techniques. In the context of an attack scenario, 25% (from total 20) of the edge clients were considered adversarial. The FL setup used a batch size of 64 for data processing during each iteration of training. The learning rate (lr) was set to 0.1, determining the step size of the model's parameter updates. Parameters like flod_threshold set at 0.5, flame_epsilon of 3000, and flame_delta at 0.001 contributed to the implementation of differential privacy in the FLAME method. Additionally, the DnC technique was configured with dnc_niters set to 5, dnc_c at 1, and dnc_b with a value of 2000, influencing the number of iterations, filtering fraction, and dimension of subsamples, respectively, within the FL framework.

Table 1 demonstrates the performance of different secure FL techniques in the presence of adversarial attacks. We noted the final accuracy after 200 rounds of iterations. Each secure aggregation technique runs with the same hyperparameters both with and without malicious attacks. Initially, we evaluate the performance of aggregation techniques without malicious attacks by selecting the highest accuracy as the benchmark for the best model. Most threat models aim to compromise the performance of the global model in its primary task. Most notably, Flod achieves the highest accuracy of 99.91%, closely followed by FedAvg, SignGuard, and DnC at 99.74%, 99.67%, and 99.53%, respectively.

Table 1: We follow the same convention of SAFEFL (Gehlhar et al. 2023) and use bold font for the best performance in each attack (represented in each column of the table); further we examine the same set of aggregation/attack models as SAFEFL for the transportation use-case.

| Attack / Aggregation | No | Trim | Krum | FLTrust | Min-Max | Min-Sum | Label flipping | Scaling |
|---|---|---|---|---|---|---|---|---|
| M1: FedAvg | 99.74% | 99.61% | **99.80%** | 99.76% | 97.97% | 99.58% | 99.65% | 51.30% |
| M2: FLTrust | 98.60% | 98.71% | 98.53% | 98.56% | 98.41% | 98.65% | 98.34% | 98.18% |
| M3: Krum | 98.38% | 98.37% | 84.31% | 97.76% | 98.52% | 98.56% | 98.61% | 98.64% |
| M4: Trim mean | 99.45% | 98.26% | 99.02% | 98.65% | 98.36% | 98.92% | 99.67% | 99.44% |
| M5: Trim Median | 99.17% | 98.00% | 98.92% | 98.11% | 98.07% | 98.66% | 99.39% | 98.99% |
| M6: FLAME | 99.41% | 99.39% | 99.26% | 99.18% | 99.27% | 99.05% | 99.52% | 99.44% |
| M7: ShieldFL | 81.44% | 79.15% | 02.92% | 43.78% | 74.61% | 82.27% | 73.03% | 66.22% |
| M8: FLOD | **99.91%** | **99.72%** | 99.63% | **99.97%** | 96.25% | **100.0%** | 98.50% | 99.85% |
| M9: DnC | 99.53% | 99.56% | 99.32% | 99.00% | 99.32% | 99.06% | 99.73% | 99.62% |
| M10: FoolsGold | 51.20% | 45.88% | 45.88% | 45.88% | 51.20% | 96.60% | 51.20% | 97.16% |
| M11: CONTRA | 51.20% | 51.20% | 02.92% | 00.00% | 51.20% | 46.14% | 51.19% | 51.20% |
| M12: SignGuard | 99.67% | 99.07% | 99.45% | 99.37% | **99.73%** | 99.39% | **99.74%** | **99.73%** |
| M13: FLARE | 98.87% | 98.23% | 98.87% | 98.89% | 96.03% | 98.96% | 99.57% | 96.62% |
| M14: Romoa | 99.21% | 98.79% | 99.47% | 99.24% | 98.79% | 99.33% | 99.63% | 99.26% |

Other methods such as Trim Mean (99.45%), FLAME (99.41%), Trim Median (99.17%), and Romoa (99.21%) also exhibit high accuracy rates above 99%. However, ShieldFL (81.44%) and techniques like FoolsGold and Contra (both at 51.20%) depict notably lower accuracies, indicating potential weaknesses or limitations in their performance in FL settings. FedAvg proves robust against krum attacks, achieving accuracies of 99.80%. Flod offers robust security against Trim, FLTrust, and Min-Sum attacks, reporting 99.72%, 99.97% and 100% accuracy, respectively. SignGuard emerges as another robust aggregation technique, securing against Min-Max, label flipping, and scaling attacks. However, FLTrust, FLAME, DnC, and SignGuard exhibit promising capabilities in mitigating poisoning attacks. FLTrust achieved a baseline aggregation performance of 98.60%, with a maximum deviation of 0.42% observed under various attack scenarios. Similarly, FLAME demonstrated robustness with an initial aggregation performance of 99.41%, experiencing a maximum deviation of 0.36% in the presence of different attacks. DnC showcased resilience with an aggregation performance of 99.53%, revealing a maximum difference of 0.53% amidst varying attack scenarios. SignGuard, another method employed in FL, attained a noteworthy aggregation performance of 99.67%, displaying a maximum difference of 0.60% under diverse attack conditions. These findings underscore the effectiveness of these techniques in maintaining robust FL models, even in the presence of adversarial attacks. The Krum attack completely compromises the secure performance of ShieldFL and Contra. Meanwhile, the FLTrust attack completely breaks the system performance of Contra. Moreover, the Trim mean and median perform well, with the maximum accuracy loss of 1.19% and 1.17%, respectively, under attack scenarios. However, other methods generally show only moderate security against poisoning attacks, with a few performing poorly against specific threat models.

After analyzing the results, we identified FLTrust, FLAME, DnC, FLOD, and Signguard as the most robust techniques in terms of privacy preservation in FL. According to SAFEFL, FLTrust is the most MPC-friendly approach. Consequently, we conducted a performance evaluation of FLTrust with MPC using various protocols including semi2k, spdz2k, replicated2k, and psReplicated2k to make the system resilient against inference attacks. Additionally, we compared this approach with MPC-based FedAvg as a benchmark model for comparison. Both take longer period for convergence of the model accuracy. In the replicated2k protocol, FedAvg demonstrates efficient performance with a quick running time of 4.049 hours for 50 iterations. Throughout the process, it engages in 216,266 communications. On the other hand, FLTrust lags behind, taking approximately 13.99 hours to complete the task, which is approximately 3.45 times longer than FedAvg. FLtrust also requires a significantly higher number of communications, totaling 3,019,221. This highlights a notable communication overhead of FLTrust, which is approximately 13.96 times higher than that of FedAvg. However, other protocols such as semi2k, spdz2k, psReplicated2k takes larger time and communication compared to replicated2k protocol.

**CONCLUSION**

The empirical study evaluates secure FL techniques within AV applications, emphasizing cyber-attack-resilient aggregation methods during training and inference phases. It builds on the evaluated/proposed techniques by (Gehlhar et al. 2023) and examines those methods on a transportation-related dataset. The results highlight varying accuracies among aggregation techniques, particularly with FLOD and SignGuard demonstrating high accuracy against diverse attacks. While some methods offer moderate security against poisoning attacks, others provide inadequate protection. Notably, FLTrust, FLAME, DnC, FLOD, and SignGuard ensure data privacy and model parameter protection during FL training for AV application development. This study significantly enhances the understanding of secure FL in AV scenarios and establishes a foundational framework for resilient future AV applications against cyber threats.

**ACKNOWLEDGEMENT**

This work is based upon the work supported by the National Center for Transportation Cybersecurity and Resiliency (TraCR) (a U.S. Department of Transportation National University

Transportation Center) headquartered at Clemson University, Clemson, South Carolina, USA. Any opinions, findings, conclusions, and recommendations expressed in this material are those of the author(s) and do not necessarily reflect the views of TraCR, and the U.S. Government assumes no liability for the contents or use thereof. Moreover, Md Jueal Mia would like to acknowledge support from the Knight Foundation School of Computing and Information Sciences (KFSCIS) at FIU, as well as the Graduate & Professional Student Committee (GPSC) at FIU.